# Symphony: Localizing Multiple Acoustic Sources with a Single Microphone Array

Weiguo Wang[1], Jinming Li[1], Yuan He[1], Yunhao Liu[1,2]
[1]School of Software and BNRist, Tsinghua University
[2]Department of Computer Science and Engineering, Michigan State University
{wwg18,li-jm19}@mails.tsinghua.edu.cn
heyuan@mail.tsinghua.edu.cn,yunhao@greenorbs.com

## ABSTRACT

Sound recognition is an important and popular function of smart devices. The location of sound is basic information associated with the acoustic source. Apart from sound recognition, whether the acoustic sources can be localized largely affects the capability and quality of the smart device's interactive functions. In this work, we study the problem of concurrently localizing multiple acoustic sources with a smart device (e.g., a smart speaker like Amazon Alexa). The existing approaches either can only localize a single source, or require deploying a distributed network of microphone arrays to function. Our proposal called *Symphony* is the first approach to tackle the above problem with a single microphone array. The insight behind *Symphony* is that the geometric layout of microphones on the array determines the unique relationship among signals from the same source along the same arriving path, while the source's location determines the DoAs (direction-of-arrival) of signals along different arriving paths. *Symphony* therefore includes a geometry-based filtering module to distinguish signals from different sources along different paths and a coherence-based module to identify signals from the same source. We implement *Symphony* with different types of commercial off-the-shelf microphone arrays and evaluate its performance under different settings. The results show that *Symphony* has a median localization error of 0.694m, which is 68% less than that of the state-of-the-art approach.

## 1 INTRODUCTION

Smart devices are proliferated in our daily life. Sound recognition is an important and popular function of smart devices. For example, smart speakers like Amazon Alexa [42], Google Home [44], Apple HomePod [43], and Alibaba Tmall Genie [41] with sound recognition support various attractive applications, including voice control of home appliances, Man-machine dialogue, entertainment center.

The location of sound is basic information associated with the acoustic source. With the fast development of smart home and office applications, there is an increasing need for acoustic source localization on the smart devices. Whether the acoustic sources can be localized largely affects the capability and quality of the smart device's interactive functions, which include but are not limited to the following cases: (1) The ability of localization enables a smart speaker to process voice commands with user location awareness. When the user is lying in bed and says 'Turn on the light', the smart speaker can smartly switch off the ceiling lamp and turn on the bedside lamp, if the user's (namely the acoustic source) location is provided. (2) Localizing the acoustic source enables a smart device, e.g. the smart safeguard device, to better perceive the real situation. For example, the device may remind the parents of possible danger when it hears abnormal sounds of windows or doors from the baby's room. (3) Knowing the source location helps to authenticate voice commands. Recent studies have uncovered the vulnerabilities of smart speakers against inaudible and malicious voice commands [25, 31, 38, 48]. To defend against these threats, the smart speaker can leverage the knowledge of voice location to accept only the commands that originate from the real locations.

Conventional approaches of acoustic source localization require to deploy multiple distributed microphone arrays. Based on the estimation of the source's time-difference-of-arrival (TDOA) or direction-of-arrival (DoA) at the arrays, the source can be localized via triangulation [5, 6, 16, 19, 27, 40, 45]. However, those solutions cannot be applied to localization with a device like the smart speaker, which is usually equipped with only a single array.

Acoustic source localization with a single array is a non-trivial problem. Note that the typical size of a microphone array is several centimeters at most, which is negligible with respect to the distance between the source and the array. As a result, the acoustic signal's propagation rays to the microphones are regarded to be parallel with each other. Due to limited spatial resolution (array size or aperture) and temporal resolution (sampling rate of the microphone), a commercial array can't separate DoAs of parallel rays. That is the so-called far-field effect.

Exploiting the multi-path propagation paves a way to tackle the above problem, however, only in the scenario of localizing a single source. In addition to the Line-of-Sight (LOS) path, VoLoc [35] leverages an additional arriving path by exploiting the nearby wall reflection (denoted as ECHO), and then localizes the far-field source after estimating DoAs of LOS and ECHO.

It's worth noticing that VoLoc assumes there is only a single source in the sound field, which largely restricts its applicability in the real world. Usually, there are multiple acoustic sources in a real scenario. For example, in the home environment, there may be families' voices, television, washing machine, and microwave oven. These sources, including the voice commands, will interfere with each other, making it very difficult for VoLoc to localize the voice.

The ability of concurrently localizing multiple acoustic sources is quite appealing. First, the localization ability can now be applied to more practical environments. A smart speaker with such ability can naturally deal with inter-source interference. In this case, interfering sounds can be intentionally treated as acoustic sources and localized as well. What's more, a smart speaker is able to interact with multiple users simultaneously, as long as they locate at different positions. Generally, the perceptual ability of a smart speaker can be further improved if the locations of multiple sources are known. For example, when the user, who is sitting on a couch and watching the television, says "Turn off the light", the smart speaker will infer that the user wants a better visual experience and then switch lights to Home Theater mode [3, 13], if the voice command at the couch is localized and the other active source at the television cabinet is also detected.

Localizing multiple acoustic sources with a single array is indeed a daunting task, with the following critical challenges: (1) The signals received at the array are a mixture of signals from multiple sources along different paths, making it extremely difficult to extract clean signals from any source. (2) The interference among the received signals blurs the relationship between signals and sources. It becomes very difficult to associate the DoA of a signal to its corresponding source, even if a DoA can be estimated. (3) The arriving paths of different sounds are diverse and unpredictable. The signals along those paths arrive after unpredictable delays. The exact arriving order of paths is unknown, thus hindering the discrimination of LOS and ECHO.

We in this paper propose *Symphony*, the first approach to localize multiple acoustic sources using a single microphone array. The design of *Symphony* stems from the following insights. (1) Although the propagation process of each source in the whole sound field is unpredictable, a microphone array will receive the signals of their arriving paths with predictable geometric patterns, determined by the prior-known layout of the array. (2) The coherence of sources exists not only among signals received by different microphones, but also among different arriving paths, LOS and ECHO.

*Symphony* exploits the redundancy in multiple microphone pairs with diverse spacing or (and) orientations to collaboratively estimate the DoAs of arriving paths. We design a novel algorithm that leverages intrinsic coherence among homologous paths to check whether two DoAs correspond to the same source. We find an interesting fact: the DoAs of LOS and ECHO should meet a certain criterion such that LOS and ECHO can intersect at a certain location. We formulate this criterion and apply it to discriminate LOS and ECHO. Our contributions are summarized as follows:

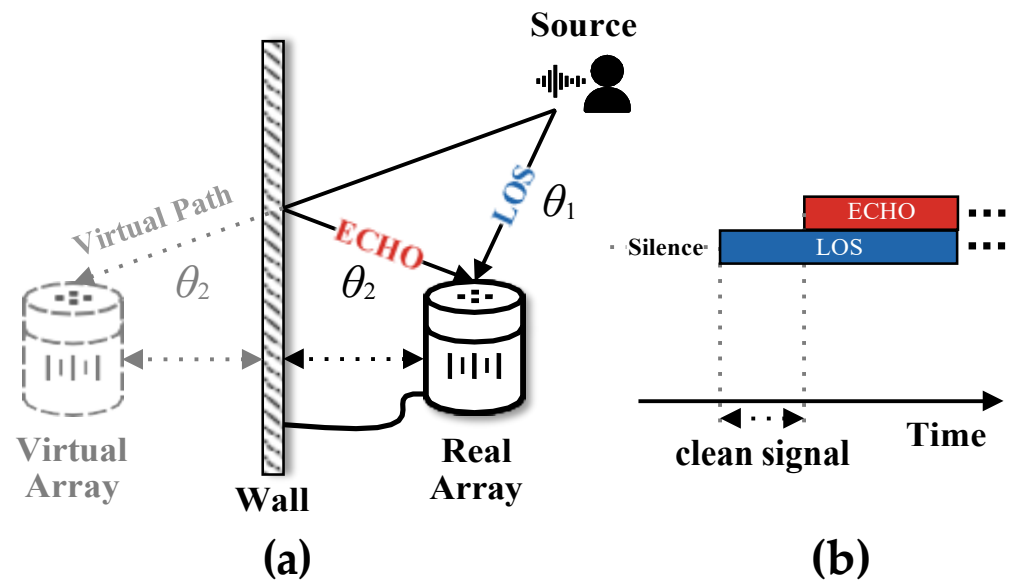

**Figure 1: (a) By viewing the nearby wall as a mirror, we can create an additional virtual array. This virtual array will receive the far-field source with DoA $\theta_2$, which is different from LOS's DoA, $\theta_1$. (b) Before the arrival of ECHO, there will be a short window of the clean signal.**

- To the best of our knowledge, *Symphony* is the first approach to localize multiple acoustic sources with a single array. The novel layout-aware design is inspired by the insight on the geometric redundancy residing in the microphone array, which is effective in resolving ambiguity induced by multiple sources.
- Symphony is a complete solution that addresses a series of technical challenges in multi-source localization. We estimate the DoAs of each path through a curve-fitting optimization process. To map DoAs to each source, we design a novel algorithm that exploits the coherence among homologous paths. We then formulate an intersection criterion to discriminate LOS and ECHO.
- We implement *Symphony* with two types of commercial off-the-shelf microphone arrays and evaluate its performance under different settings. The results show that *Symphony* has a median localization error of 0.694m, which is 68% less than that of VoLoc.

**Roadmap.** § 2 introduces the background knowledge. § 3 presents the overview. § 4 introduces the propagation model of multiple sources. In § 5, § 6, and § 7, we elaborate on the design of *Symphony*. § 8 tackles practical issues. § 9 presents the evaluation results. § 10 discusses the related work. § 11 and § 12 respectively discuss and conclude this work.

## 2 PRIMER

**Far-Field Effect.** The smart speaker typically holds a single microphone array. There is a critical barrier for a single microphone array to localize the source: the far-field effect. A source is considered to be in the far-field if [7]

$$L \geq \frac{2d^2}{\lambda} \tag{1}$$

where $L$ is the distance between the source and the array, $\lambda$ is the wavelength of the arriving wave, and $d$ is the inter-distance between two microphones. In practice, a voice source naturally meets such condition. The fundamental frequency of the human speech (except singing) is typically less than 500 Hz, and the corresponding $\lambda >$ 0.66m. For an array with size 15cm, the source can be viewed as in the far field as long as it is 6.8cm away from the source. That is why the far-field effect is leveraged in microphone array processing for many applications such as [4, 9, 10, 14, 30].

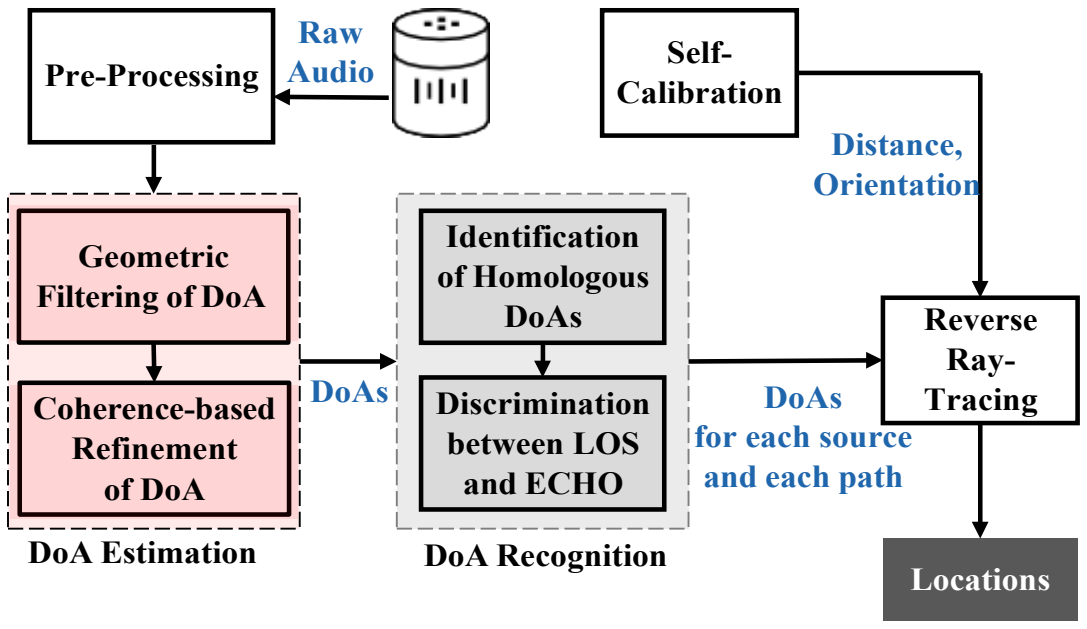

**Figure 2: *Symphony* System Overview.**

However, the far-field effect is not friendly to the localization task using a single array. Intuitively, if we can precisely obtain the DoAs of each propagation rays, the source location can be easily localized as the intersection of these rays. Unfortunately, for the far-field sources, the propagation rays are nearly parallel, the DoAs of these rays are too close to be separated due to limited spatial resolution.

**Localization with a Single Array.** VoLoc [35] leverages the nearby wall reflection to tackle the far-field effect, enabling a single array to localize the far-field source. We introduce this idea from an interesting perspective: viewing the nearby wall as a mirror and creating a virtual array. Fig. 1(a) demonstrates the idea. The smart speaker is typically placed near a wall for power supply. If we view the wall as a mirror, a virtual array will appear behind this 'mirror' according to the plane mirror imaging principle. In other words, besides the real array, we create an additional virtual array. The far-field source will arrive at the real and the virtual array by the LOS path and the virtual path. Due to the relatively large distance between these two arrays, the LOS path and the virtual path will no longer be parallel, but have two distinguishable DoAs $\theta_1$ and $\theta_2$. This means we have the opportunity to resolve DoAs of LOS and ECHO, and localize the far-field source via reverse ray-tracing.

To estimate DoAs of LOS and ECHO, VoLoc takes advantage of pauses in a voice command. A voice command is usually preceded by silence. This means they will be a short time window during which the LOS signal is clean, as shown in Fig. 1(b). The clean signal can then be used to directly derive the DoA of LOS. Meanwhile, since the ECHO path is a delayed version of the LOS path, VoLoc views the clean signal as a template to model and cancel ECHO with appropriate alignment.

In practice, there are many other sources in a home, such as other families' voices, television, washing machine, microwave oven. These sources will certainly interfere with voice commands in many scenarios, and largely restrict the application of VoLoc in the real world. The immediate effect of the inter-source interference is that the VoLoc's assumption of pauses or silences may not be valid. It is unlikely to obtain the clean signal of the targeted source. What's more, the microphones receive a mixture of signals from multiple sources along the LOS and ECHO paths, making it difficult to model and cancel other arriving paths.

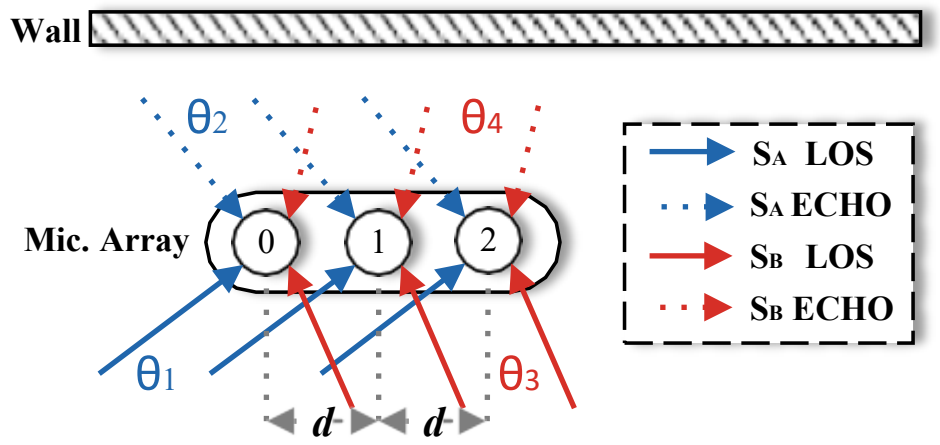

**Figure 3: The propagation model of two sources, $S_A$ and $S_B$.**

## 3 *SYMPHONY* OVERVIEW

To tolerate such inter-source interference and, more importantly, to improve the perceptual ability of a smart speaker, we propose a novel localization approach, *Symphony*, which enables a smart speaker to localize multiple sources concurrently.

Fig. 2 shows the architecture of our system. To reverse ray-trace and localize multiple sources (§ 7), *Symphony* needs to obtain the DoA of LOS and ECHO for each source, as well as the array's relative position to the nearby wall. *Symphony* adopts a two-stage scheme to obtain DoAs. (1) DoA Estimation (§ 5): After pre-processing the raw audio, *Symphony* uses geometric redundancy to produce high-resolution DoAs and leverages signal coherence to refine DoA results. (2) DoA Recognition (§ 6): *Symphony* then figures out which DoAs belong to the paths coming from the same source, and discriminates between LOS and ECHO. To self-calibrate its own location, *Symphony* transmits probing pulses (§ 8) to measure the distance and orientation to the nearby wall.

## 4 PROPAGATION MODEL

We begin with building the propagation model of multiple sources with multiple paths. The model mainly focuses on two paths: LOS and ECHO. This model will guide us in the estimation and the recognition of DoAs later.

Suppose two sources, $S_A$ and $S_B$, are active simultaneously in the room, and a linear microphone array with inter-distance $d$ is used to record signals. In Fig. 3, two main paths of each source are illustrated: the direct path (LOS) and the wall reflection path (ECHO). The Source $S_A$ arrives at the array with DoA $\theta_1$ directly, and its ECHO path arrives with DoA $\theta_2$. If we choose the first microphone $M_0$ as the reference, the signal received by the microphone $M_n$ at time $t$ can be expressed as follows:

$$
\begin{aligned}
y_n(t) = \; & a_A^{los} S_A\left(t - \tau_A^{los} - \mathrm{F}_n(\theta_1)\right) + a_A^{echo} S_A\left(t - \tau_A^{echo} - \mathrm{F}_n(\theta_2)\right) \\
& + a_B^{los} S_B\left(t - \tau_B^{los} - \mathrm{F}_n(\theta_3)\right) + a_B^{echo} S_B\left(t - \tau_B^{echo} - \mathrm{F}_n(\theta_4)\right) \\
& + \dots \qquad n = 0, 1, 2, \dots,
\end{aligned}
\tag{2}
$$

where $a$ is the attenuation factor, and $\tau_{\bullet}^{los}$ and $\tau_{\bullet}^{echo}$ are the propagation times from the source $S_{\bullet}$ to reference microphone $M_0$ by LOS and ECHO, respectively, and $\mathrm{F}_n(\theta)$ denotes the relative delay between microphone $M_0$ and $M_n$ for the signal coming from DoA $\theta$. For a uniform linear array with inter-distance $d$,

$$
\mathrm{F}_n(\theta) = n \frac{d \cos\theta}{v} \tag{3}
$$

where $v$ is the speed of sound. For simplicity, only the LOS path and the ECHO path are considered in Eq. 2, and other paths are not included. We discuss the reason at the end of this section.

One of the most popular methods to estimate DoA of wideband signals, GCC-PHAT [23], is based on generalized cross-correlation. Consider two signals received by microphone $M_n$ and $M_m$, the cross-correlation function (CCF) between $y_n$ and $y_m$ is defined as

$$Cor_{n,m}(\tau) = E\left[y_n(t-\tau)\, y_m(t)\right] \quad (4)$$

In the single-source case in a free-space environment (no multipath effect), there is only one main correlation peak, and the time shift of the peak $\tau^* = \arg\max_\tau Cor_{n,m}(\tau)$ captures the relative delay of the source to the microphone pair $\langle n, m\rangle$, which can then be used to calculate DoA according to above Eq. 3.

The above method is inappropriate in the case of multiple sources with multiple paths. To localize these sources, *Symphony* needs to estimate DoAs of multiple paths of multiple sources, especially LOS and ECHO. Intuitively, instead of determining only one maximum correlation peak, we may determine multiple peaks to estimate multiple DoAs. Substituting Eq. 2 into Eq. 4 and assuming sources are mutually uncorrelated, we can find that there are a series of correlation peaks. Table. 1 lists these peaks' time shifts.

| Peak Type | Source | |
|---|---|---|
| | $S_A$ | $S_B$ |
| LOS-LOS | $\frac{d}{v}(m-n)\cos\theta_1$ | $\frac{d}{v}(m-n)\cos\theta_3$ |
| ECHO-ECHO | $\frac{d}{v}(m-n)\cos\theta_2$ | $\frac{d}{v}(m-n)\cos\theta_4$ |
| LOS-ECHO | $\frac{d}{v}(m\cos\theta_1 - n\cos\theta_2) + \tau_A^{los} - \tau_A^{echo}$ | $\frac{d}{v}(m\cos\theta_3 - n\cos\theta_4) + \tau_B^{los} - \tau_B^{echo}$ |
| ECHO-LOS | $\frac{d}{v}(m\cos\theta_2 - n\cos\theta_1) + \tau_A^{echo} - \tau_A^{los}$ | $\frac{d}{v}(m\cos\theta_4 - n\cos\theta_3) + \tau_B^{echo} - \tau_B^{los}$ |

**Table 1: Peaks in cross correlation between $M_n$ and $M_m$.**

Peaks in Table. 1 can be divided into two categories: **pure peaks** (LOS-LOS or ECHO-ECHO) and **hybrid peaks** (LOS-ECHO or ECHO-LOS). Such division is actually based on two basic facts:

- Firstly, let's zoom in on the array to observe the signal propagation on *a certain path*. We find that this path arrives at each microphone after different but short delays. These short delays are just captured by pure peaks. As shown in Table 1, the time shifts of pure peaks depend only on the DoAs of each path. This dependency implies that as long as we identify all pure peaks, we can estimate the DoA of each path. We elaborate on this in § 5.
- Secondly, let's zoom out to the whole sound field to observe the signal propagation from *a certain source*. We can notice that this source has two main arriving paths, LOS and ECHO. Shortly after LOS arriving at the array, ECHO also reaches the array. The delay between LOS and ECHO is actually captured by hybrid peaks. Here, we point out that hybrid peaks will act as a bridge between the paths from the same source, which can be used to recognize the DoAs of their sources. We elaborate on this in § 6.

In summary, *Symphony* estimates DoAs based on pure peaks, and recognizes source for each DoA based on hybrid peaks. Before introducing our design, we discuss the following question.

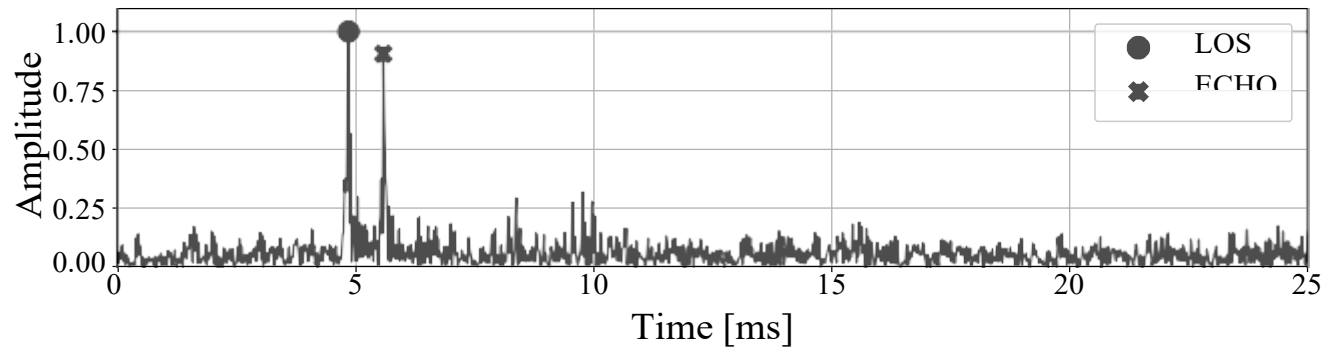

**Figure 4: Channel impulse response. Both LOS and ECHO are overwhelmingly stronger than following arriving paths.**

■ **The model considers only two paths, LOS and ECHO. What about the latter arriving paths?** LOS is the first-arrival and strongest path because it traverses the shortest distance. Meanwhile, ECHO, which comes from the nearby wall reflection, traverses only a slightly longer distance than LOS and thus experiences nearly the same propagation attenuation as LOS[1]. On the other hand, the latter paths arrive after traversing much longer distances. This means the latter paths are considerably weaker than LOS and ECHO. We validate this by measuring channel impulse response (CIR) in a living room. We place the array 30cm away from the wall. As shown in Fig. 4, there are two distinct spikes with nearly the same amplitudes. These two spikes exactly correspond to LOS and ECHO. Their energy is dominant over all other paths. This explains why it is reasonable to consider only LOS and ECHO.

## 5 DOA ESTIMATION

As we mentioned in § 4, DoAs can be calculated by the time shifts of pure peaks. In this section, we introduce how to identify all pure peaks among a series of correlation peaks. In § 5.1, we take advantage of the geometric redundancy of multiple microphone pairs to filter out undesired peaks. However, this approach is not completely reliable due to the limited spatial resolution. Therefore, in § 5.2, we further leverage the intrinsic coherence among signals to identify all pure peaks.

### 5.1 Geometry-Based Filtering of DoA

We now introduce our method to identify pure peaks (LOS-LOS and ECHO-ECHO) of multiple sources. Our insight is that instead of using a single microphone pair, we can take advantage of redundancy among multiple microphone pairs with different layouts. Without loss of generality, we first use a uniform linear array to explain and validate our idea. Then we extend the idea to other layouts.

*5.1.1 Uniform Linear Array (ULA).* We observe that for microphone pair $\langle n, m\rangle$, the time shifts of pure peaks are directly proportional to the subtraction of two microphone serial numbers, $m-n$. If we revisit the cell ⟨LOS-LOS, $S_A$⟩ in Table 1, we will find that the relation between the time shift $\tau$ and the variable $m-n$ is a linear function: $\tau = k(m-n)$, where the slope $k = \frac{d}{c}\cos$. This linear relation only holds for pure peaks, but not for other peaks. Therefore, such relation can be exploited to identify pure peaks.

We conduct the following proof-of-concept experiment to validate our idea. In a living room, we let two speakers simultaneously play two recorded voice commands (denoted as $S_A$ and $S_B$) at different places. A uniform linear array with 4 microphones is placed 30 cm away from the wall to record signals. Fig. 5(a) shows the cross-correlation functions between three microphone pairs $\langle 0,1\rangle$, $\langle 0,2\rangle$ and $\langle 0,3\rangle$. In these plots, the markers with the shape of triangle denote the local maximums of correlation, which are the candidates of pure peaks. The markers with the shape of circular and cross are the ground truths of pure peaks. The ground truths are obtained by inserting preambles (prior known signals) immediately proceeding the voice commands. Fig. 5 (a) emphasizes the time shifts of pure peaks for LOS $S_A$. As we can see, the time shifts of these pure peaks change linearly with the subtraction of the microphone serial number, $m-n$.

[1] The energy loss resulting from wall reflection is also negligible. Typically, more than 95% signal energy remains after reflection due to highly impedance mismatch between the air and the wall.

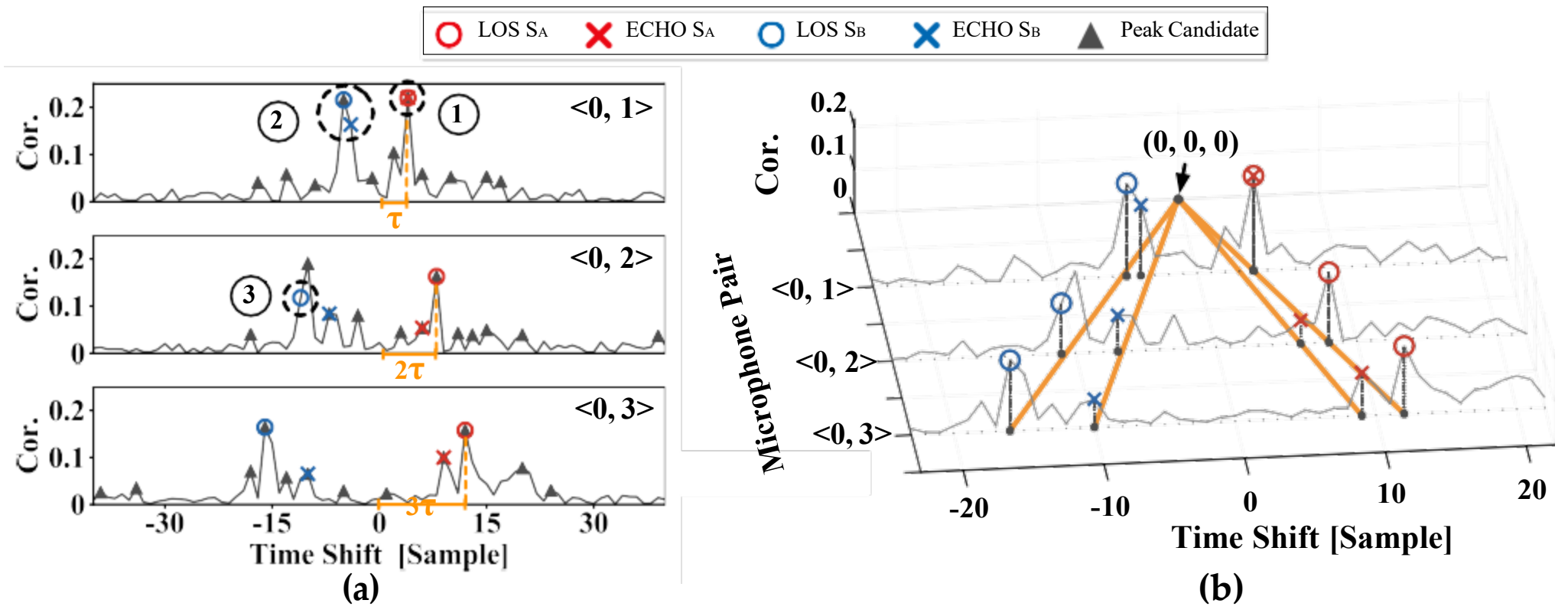

**Figure 5: (a) Cross Correlation between microphone pairs $\langle 0,1\rangle$, $\langle 0,2\rangle$ and $\langle 0,3\rangle$. (b) We observe that pure peaks across three microphone pairs can nearly fit a straight line passing the origin $(0,0,0)$.**

To be more clear, we incorporate these three cross-correlation functions into the same coordinate system, as shown in Fig. 5 (b). It is very interesting to see that, if we sequentially connect the markers of the ground truths that correspond to the same path, the pure peaks can nearly form a straight line passing the origin $(0,0,0)$. This phenomenon motivates us to exploit such linear relation to find pure peaks. However, before introducing our method, it is worthwhile to analyze the following problems.

- **Problem 1: Peak Overlap.** Some markers of the ground truths are so close that it is difficult to separate them (e.g., ① in Fig. 5 (a)), or the marker is no longer the local maximum because of being suppressed by the adjacent marker (e.g., ②).
- **Problem 2: Peak deviation.** As shown by ③, some markers of the ground truths fail to locate at the local maximum, and instead is one-sampling-point away from the nearest peak.

In fact, both problems are caused by limited spatial resolution. Recall that the value we can measure is the time shift of peak, and the value we intend to obtain is the DoA $\theta$. However, the sampling rate of the array $F_s$ will limit the resolution of measurable time shift. The DoA $\theta \in [0, \pi]$, which is continuous, is mapped into $\lfloor \frac{(m-n)d}{c} F_s \rfloor$ discrete bins. Such mapping may introduce an additional conversion error, thus introducing the deviation between the ground truth and the peak candidates, such as ③. If two DoAs are so close that fall into the same discrete bin, the array is impossible to separate them, such as ①. On the other hand, as $m-n$ increases, the number of discrete bins also increases, thus providing higher resolution. This explains why the markers of $S_A$ no longer stay in the same bin in pair $\langle 0,2\rangle$, and also explains why Problems 1 and 2 are unlikely to occur in pair $\langle 0,3\rangle$.

| Term | Brief description |
|---|---|
| $P_{n,m}$ | The set of time shifts of peak candidates in pair $\langle n,m\rangle$. |
| $\tau_{n,m}$ | The time shift of peak candidate and $\tau_{n,m} \in P_{n,m}$. |
| $c_i$ | One possible combination of peak candidates across multiple pairs. In 4-mic ULA case, $c_i \in P_{0,1} \times P_{0,2} \times P_{0,3}$. |
| $w_{n,m}$ | The penalty factor of pair $\langle n,m\rangle$ and equals to $\lvert m-n\rvert$. |

**Table 2: Definition of Terminology.**

The analysis above suggests that the occasional errors are inevitable, and then it is unrealistic to apply a strict criterion to directly identify pure peaks according to the linear relation. To tolerate such imperfections, we formulate DoAs estimation into a curve-fitting problem using a fitting metric. This metric evaluates how well peak candidates across each pair fit a curve. In particular, the metric used for the case of ULA is expressed as follows (Table 2 defines the terminology):

$$L(c_i) = \min_k \frac{1}{N-1} \sum_{\tau_{n,m} \in c_i} w_{n,m} \left( k(m-n) - \tau_{n,m} \right)^2 \quad (5)$$

Intuitively, we select one candidate peak from each peak set for each microphone pair ($P_{n,m}$), and each possible selection constitutes a combination $c_i$. The whole selection space is the Cartesian product of the peak sets of pairs ($P_{0,1} \times P_{0,2} \times P_{0,3}$). The metric $L(c_i)$ evaluates the shortest distance between candidate peaks of $c_i$ and the regression line $y = k(m-n)$. The $c_i$ with the smaller $L(c_i)$ is more likely to be the combination of pure peaks. As we discussed before, as $m-n$ increases, the resolution of pair $\langle n,m\rangle$ also increases, which means the pair with larger $m-n$ tends to have smaller error variance. This is a classic heteroscedasticity problem [11]. Following the idea of weighted least squares, we assign different pairs with different penalty factors $w_{n,m}$ inversely proportional to their error variances. In this way, we augment results from pairs with different error variances, and can estimate DoA more accurately.

| $i$ | $c_i$ [sample] | L($c_i$) | $k^*$ | pure peak | source |
|---|---|---|---|---|---|
| 1 | (4, 8, 12) | 0.00 | 4.00 | Yes | $S_A$ |
| 2 | (6, 13, 20) | 0.191 | 6.633 | | |
| 3 | (-1, -3, -5) | 0.191 | -1.633 | | |
| 4 | (-4, -7, -10) | 0.191 | -3.367 | Yes | $S_B$ |
| 5 | (-5, -10, -16) | 0.258 | -5.276 | Yes | $S_B$ |
| 6 | (2, 6, 9) | 0.328 | 2.990 | Yes | $S_A$ |
| 7 | (2, 6, 7) | 1.004 | 2.439 | | |
| 8 | (-4, -10, -13) | 1.004 | -4.439 | | |

**Table 3: The ranking list of $c_i$ using metric L. (Due to the space issue, only top-8 entries are listed)**

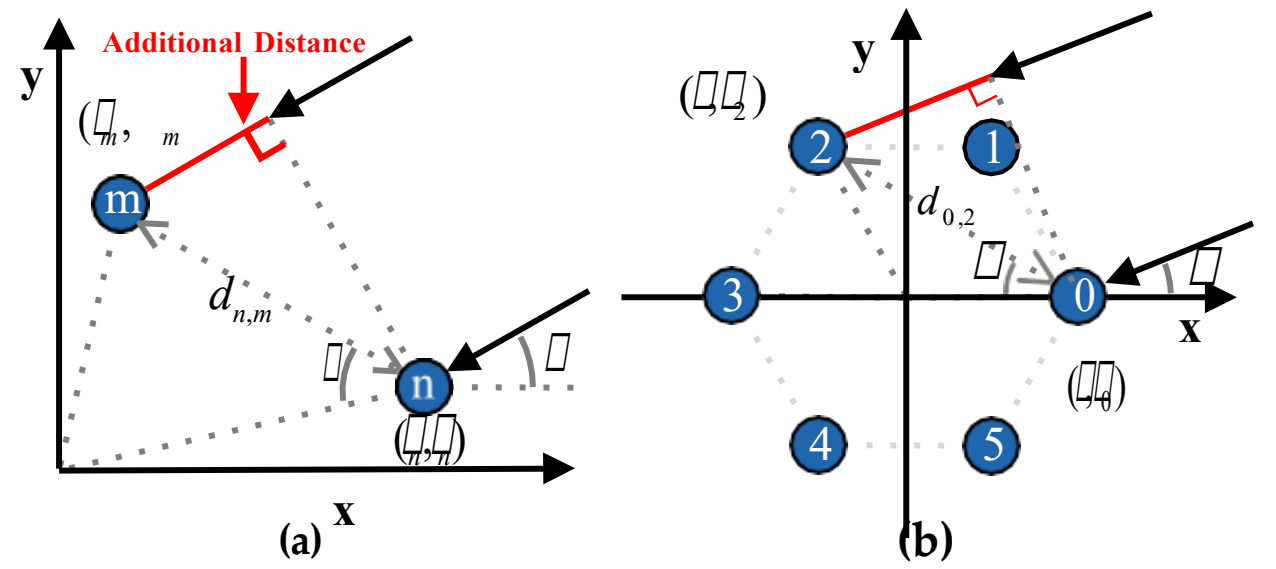

**Figure 6: (a) two mics. in a 2D space. (b) 6-mic circular array.**

We apply this metric to the previous proof-of-concept experiment, by calculating L of each $c_i$ in Fig. 5. Table 3 ranks each $c_i$ in the ascending order of L($c_i$). The slopes $k^*$ of the lines fitted by $c_i$ are also included. Based on the ground truths in Fig. 5, we mark the entries that belong to pure peaks. We can see that pure-peak entries get high ranks: first, fourth, fifth, and sixth, respectively. However, some entries not corresponding to pure peaks also get relatively high ranks: second and third, which may mislead the identification of pure peaks. The results show that this method can eliminate many ambiguities by ruling out low-ranking entries, but is incapable of identifying pure peaks confidently. To tackle this problem, we will refine the ranking results to finally determine DoAs in 5.2.

*5.1.2 Extend to General Geometry.* We extend our idea to other arrays with different layouts. The main difference is that the curve of pure peaks is no longer a straight line, but a more complicated curve. Fortunately, because the geometry of each array is prior-known and well defined, the exact mathematical expression of this curve can be derived.

Let's first consider two microphones $M_n$ and $M_m$ on a plane whose polar coordinates are $(\rho_n, \beta_n)$ and $(\rho_m, \beta_m)$, as shown in Fig. 6 (a). When there is the signal of a path arrives at this pair with DoA $\theta$, the relative delay between $M_n$ and $M_m$ can be calculated as follows

$$F[\langle n,m\rangle,\theta] = \frac{\Delta d}{v} = \frac{d_{n,m}\cos(\pi - (\gamma - \beta_n) - \theta)}{v} \tag{6}$$

where $\Delta d$ denotes the additional propagation distance, and $d_{n,m}$ is the distance between $M_n$ and $M_m$, and $\gamma$ is a fixed angle determined by the coordinates of $M_n$ and $M_m$. Again, because the geometry is given, the above parameters are all fixed and known. This actually reveals a basic fact that the relative delay F only depends on microphone pair $\langle n, m\rangle$, and DoA $\theta$. As a concrete example, we apply F $[\langle n, m\rangle, \theta]$ to another popular array: 6-mic circular array, as shown in Fig. 6 (b). The polar coordinate of $M_n$ is $(\rho, n\beta)$. The relative delay can be rewritten as follows

$$F[\langle n,m\rangle,\theta] = \frac{\Delta d}{v} = \frac{2\rho\sin((m-n)\beta/2)\cos(\pi/2 - (m+n)\beta/2 - \theta)}{v} \tag{7}$$

A more general fitting metric to estimate DoAs across multiple pairs can be rewritten as follows

$$L(c_i) = \min_{\theta} \frac{1}{N-1} \sum_{\tau_{n,m}\in c_i} w_{n,m} \left(F[\langle n,m\rangle,\theta] - \tau_{n,m}\right)^2 \tag{8}$$

## 5.2 Coherence-Based Refinement of DoA

The results in Table 3 show that, the metric F can eliminate many undesired combinations by ranking, but we cannot determine the pure-peak combinations confidently. To solve this problem, we will refine these ranking results to determine pure-peak combinations.

Besides the geometric redundancy, the intrinsic coherence among signals received by microphones can also be exploited to identify pure peaks. Recall that pure peaks capture a certain path's relative delays among microphones. If we can compensate for these arriving delays based on pure peaks, signals received by microphones can be coherent with respect to a certain path. Based on this fact, we refine each entry $c_i$ by checking whether the time shifts of $c_i$ can make signals coherent with respect to a certain path. Next, a 4-mic linear array, again, is taken as an example to introduce our method. Algorithm 1 describes the refinement procedure.

**Algorithm 1:** Refinement of Pure-Peak Entries

1 **Input:** four signals received by 4-mic, $y_0, y_1, y_2$ and $y_3$.
2 Compute CCF $Cor_{0,3}$ between $y_0$ and $y_3$.
3 **for** *each* $c_i = (c_i^1, c_i^2, c_i^3)$ *in Table 3* **do**
4 Shift the signal $y_1$ and $y_2$ using the relative delays $c_i^1$ and $c_i^2$ to in line with $y_0$.
5 Average two shifted signals and the signal $y_0$, and denote the result as $y_{<0,1,2>}$.
6 Compute CCF $Cor_{<0,1,2>,3}$ between $y_{<0,1,2>}$ and $y_3$
7 **if** $Cor_{<0,1,2>,3}(c_i^3) - Cor_{0,3}(c_i^3) > Threshold$ **then**
8 **Identify** $c_i$ as a pure-peak combination.
9 **end**
10 **end**

In Algorithm 1, we leverage pure peaks in CCF to assert coherence. Specifically, if $c_i$ is a pure-peak combination of a certain path, after aligning and averaging the first three signals $y_0$, $y_1$ and $y_2$ (Line 4 and 5), we can accurately compensate for the arriving delays of this path and constructively enhance the path. The enhanced version of the signal is denoted by $y_{<0,1,2>}$. Theoretically, when we correlate the enhanced signal $y_{<0,1,2>}$ with another signal $y_3$, the pure peak corresponding to this path (located at $c_i^3$) in CCF $Cor_{<0,1,2>,3}$ will rise significantly, compared with the original

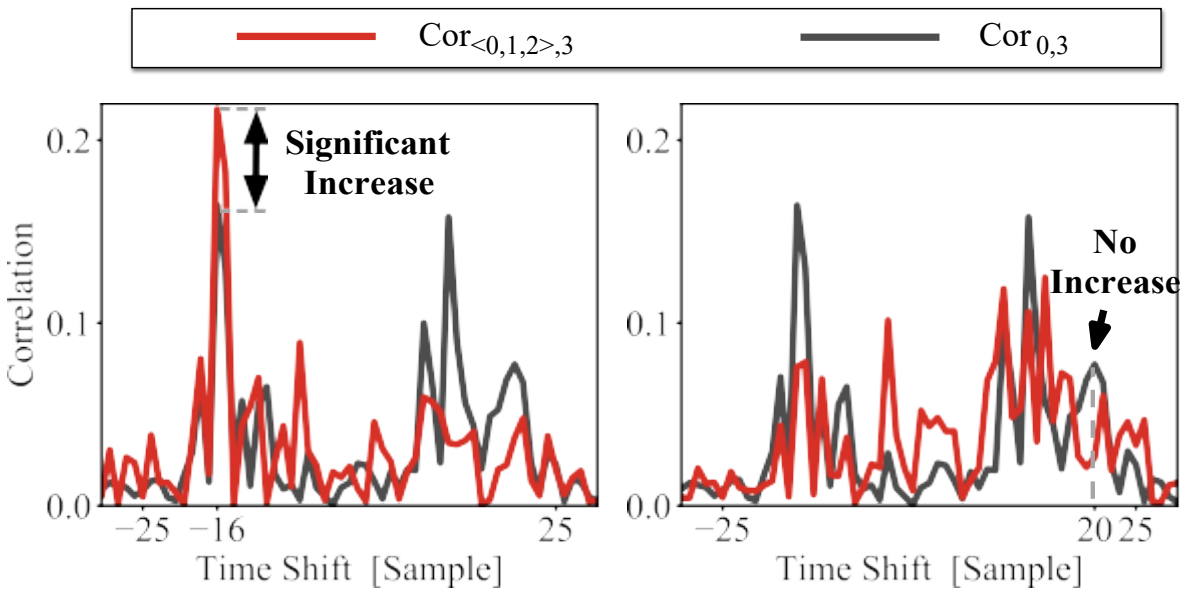

**(a)** identify $c_5$ = (-5, -10, -16) **(b)** identify $c_2$ = (6, 13, 20)

**Figure 7: Coherence-based refinement.**

CCF $Cor_{0,3}$ between $y_0$ and $y_3$. If so (Line 7), $c_i$ is identified as a pure-peak combination.

We validate this idea in the previous proof-of-concept experiment. We compute $Cor_{<0,1,2>,3}$ using the pure-peak combination $c_5$ in Table 3, and the non-pure-peak combination $c_2$. As expected in Fig.7 (a), the value of $Cor_{<0,1,2>,3}$ at $c_5^3$ (sample -16) has a significant increase, while in Fig.7 (b), no increase is observed at $c_2^3$ (sample -20). In summary, if there is an increase in $Cor_{<0,1,2>,3}$ at $c_i^3$, $c_i$ can be determined as a pure-peak combination.

**DoA Calculation.** Once we identify all pure-peak combinations, we can estimate DoAs of all paths based on the curves fitted by these combinations. To calculate DoAs based on pure peaks, we need to additionally consider the sampling rate $F_s$. For ULA, the slope $k^*$ of the line fitted by one pure-peak combination equals to $\frac{d}{v}\cos\theta \cdot F_s$. Therefore, the DoA $\theta$ can be calculated by $\arccos\frac{vk^*}{dF_s}$. For other arrays, the DoA $\theta$ corresponding to each combination has already been obtained after solving Eq. 8 (here $\tau_{n,m}$ needs to be in the form of continuous time, instead of discrete sample).

# 6 DOA RECOGNITION

In § 5, we have identified all pure-peak combinations. In this section, we firstly recognize which two pure-peak combinations belong to the same source (§ 6.1). Then in § 6.2, we distinguish their path types, i.e., which is LOS and which is ECHO.

## 6.1 Homologous Identification of DoA

Here, we identify which DoAs belong to the same source. A basic fact is that for a certain source, the ECHO path is a delayed version of the LOS path. These two paths, coming from the same source, are coherent. *Symphony* exploits hybrid peaks to capture such inter-path coherence, thus finding DoAs coming from the same source.

Without loss of generality, let's take $c_5$ in Table 3 as an example. From the ground truth, we know that $c_5$ corresponds to the LOS path of $S_B$. Similar to Algorithm 1, after aligning and averaging the first three signals based on $c_5$, we can obtain $y_{<0,1,2>}$, and then the LOS path of $S_B$ will be constructively enhanced. Note that another received signal $y_3$ actually receives the signals of the LOS and the ECHO paths of $S_B$, which both are coherent with that of the enhanced LOS path of $S_B$ in $y_{<0,1,2>}$. When we correlate $y_{<0,1,2>}$ with $y_3$, two peaks of CCF $Cor_{<0,1,2>,3}$ will increase: (1) LOS-LOS, which is the correlation between the enhanced LOS path in $y_{<0,1,2>}$ and the LOS path in $y_3$. (2) LOS-ECHO, which is the correlation between the enhanced LOS path in $y_{<0,1,2>}$ and the ECHO path in $y_3$. These two peaks are located at

$$\begin{cases} c_5^3, & \text{(LOS-LOS)} \\ c_5^3 + \tau_B^{los} - \tau_B^{echo}, & \text{(LOS-ECHO)} \end{cases} \quad (9)$$

where $c_5^3$ is the third element of $c_5$. Similarly, if we use $c_4$ in Table 3, which corresponds to the ECHO path of $S_B$, to calculate the enhanced CCF (denote this CCF as $Cor'_{<0,1,2>,3}$), we can also observe two peaks in CCF $Cor'_{<0,1,2>,3}$ will increase: ECHO-ECHO and ECHO-LOS, which are located at

$$\begin{cases} c_4^3, & \text{(ECHO-ECHO)} \\ c_4^3 + \tau_B^{echo} - \tau_B^{los}, & \text{(ECHO-LOS)} \end{cases} \quad (10)$$

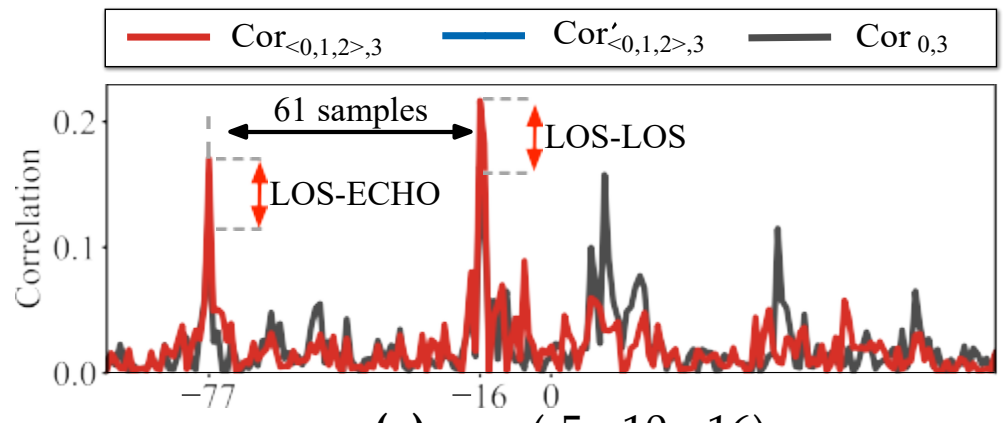

**(a)** $c_5$ = (-5, -10, -16)

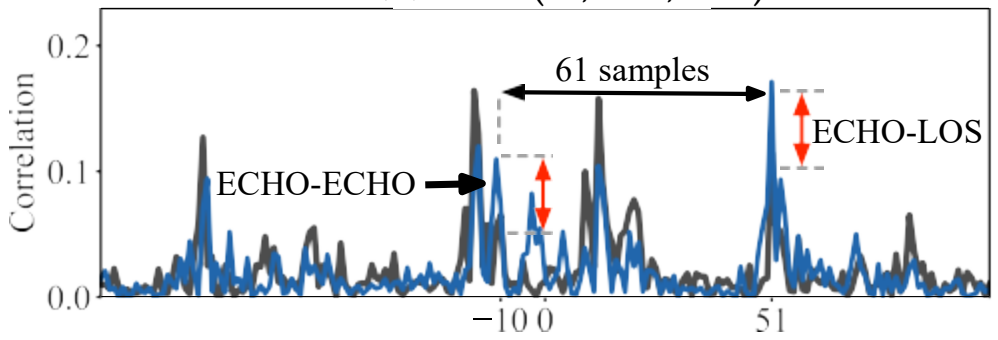

**(b)** $c_4$ = (-4, -7, -10)

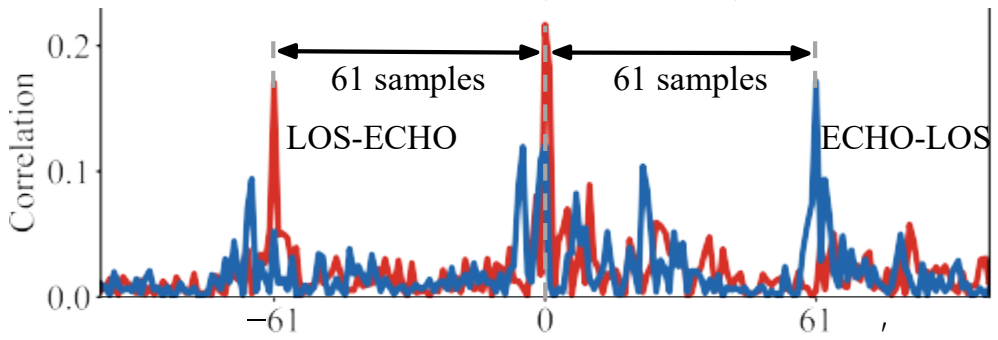

**(c)** Shifted version of $Cor_{<0,1,2>,3}$ and $Cor'_{<0,1,2>,3}$

**Figure 8: Identify whether $c_5$, $c_4$ belong to the same source.**

Now, let's compare Eq. 9 and 10. An interesting observation is that the locations of hybrid peaks that get enhanced (LOS-ECHO in Eq. 9 and ECHO-LOS in Eq. 10) are associated by a term $\tau_B^{los} - \tau_B^{echo}$. This is because the combinations $c_5$ and $c_4$ correspond to the same source $S_B$, thus both enhanced hybrid peaks capture the same arriving delay of source $S_B$ between LOS and ECHO.

Such association actually provides us an additional constraint to determine whether two pure-peak combinations $c_i$ and $c_j$ ($i \neq j$) belong to the same source. We take the following steps to apply this constraint:

(1) Fetch two CCFs $Cor_{<0,1,2>,3}$ and $Cor'_{<0,1,2>,3}$ which have already been computed in Algorithm 1 by using $c_i$ and $c_j$.
(2) Shift two CCFs $Cor_{<0,1,2>,3}$ and $Cor'_{<0,1,2>,3}$ by $c_i^3$ and $c_j^3$, respectively, and denote the results as $\overline{Cor}_{<0,1,2>,3}$ and $\overline{Cor}'_{<0,1,2>,3}$
(3) Check whether $\overline{Cor}_{<0,1,2>,3}$ and $\overline{Cor}'_{<0,1,2>,3}$ have two peaks separately, whose time shifts are symmetric about the origin and the values of the peaks get a considerable increase. If so,

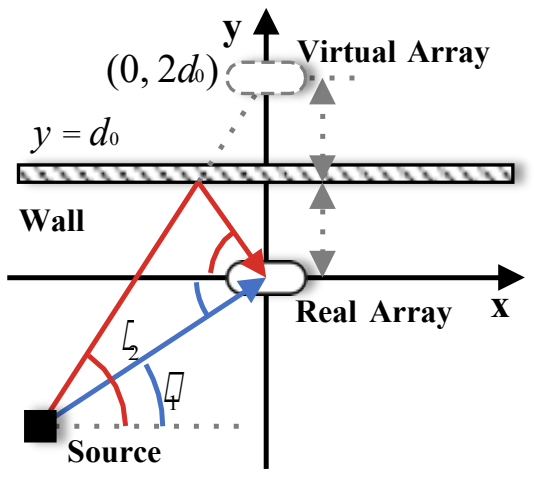

**Figure 9: To ensure that LOS can intersect with ECHO , the absolute value of the slope of LOS should be smaller than ECHO. i.e., $|\tan\theta_1| < |\tan\theta_2|$.**

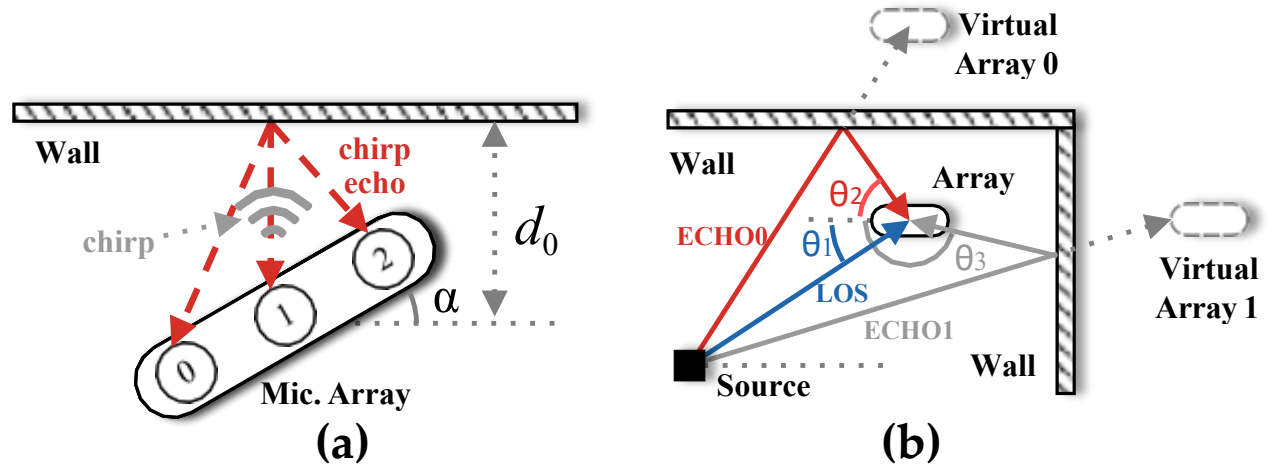

**Figure 10: (a) The distance $d_0$ and the orientation $\alpha$ to the nearby wall can be measured via the chirp pulses. (b) When the smart speak is in the corner, besides LOS, there will be two wall reflections, ECHO0 and ECHO1.**

these two combinations $c_i$ and $c_j$ are identified as belonging to the same source.

The idea is also validated in our proof-of-concept experiment. We compute two CCFs $Cor_{<0,1,2>,3}$ and $Cor'_{<0,1,2>,3}$ using $c_5$ = (-5, -10, -16) and $c_4$ = (-4, -7, -10) in Table 3 which belong to same source, $S_B$. Fig. 8 plots the results as well as the original $Cor_{0,3}$. In Fig. 8 (a), two peaks located at sample -16 and -77 increase considerably. Similarly in Fig. 8 (b), two peaks located at sample -10 and 51 also increase. We notice that the interval between two enhanced peaks in each subfigure is the same, equal to 61 samples. Therefore, once shifting CCFs $Cor_{<0,1,2>,3}$ and $Cor'_{<0,1,2>,3}$ by $c_5$ (sample -16) and $c_4$ (sample -10) respectively, we can expect that the LOS-ECHO peak locating at sample -77 in Fig. 8 (a) and the ECHO-LOS peak locating at sample 51 in Fig. 8 (b) will be symmetric about the origin (sample -61 and 61 ), as illustrated in Fig. 8 (c). In other words, by checking such symmetry, we can determine whether two combinations belong to the same source.

## 6.2 Discrimination between LOS and ECHO

Once we find out which two pure-peak combinations belong to the same source, the next is to determine which is LOS or ECHO.

Fig. 9 illustrates our observation. The blue line and the red line represent LOS and ECHO, respectively. To make sure that these two lines can intersect in Quadrant III or IV of this coordinate plane, the absolute value of the slope of the blue line should be smaller than that of the red line, namely $|\tan\theta_1| < |\tan\theta_2|$ .

Based on such observation, we propose a simple but efficient approach to distinguish LOS and ECHO. Assume we have recognized that two pure-peak combinations belong to the same source. Then, we compute their DoAs based on the fitted curves, and compare $|\tan|$ function of these two DoAs. The one with smaller $|\tan|$ is identified as LOS, and the other is ECHO.

We validate this approach by applying this criterion to $c_5$ and $c_4$ in Table 3, which are LOS and ECHO of $S_B$, separately. The $|\tan|$ values computed with $c_5$ and $c_4$ are 0.850 and 1.847. As expected, the $|\tan|$ value of LOS is smaller than that of ECHO.

# 7 LOCALIZATION VIA REVERSE RAY-TRACING

Now, the DoAs of LOS and ECHO for each source, $\theta_1$ and $\theta_2$, are available. We then localize each source via reverse ray-tracing.

We construct the coordinate system and let the array as the origin point, the nearby wall as the line $y = d_0$, as illustrated in Fig. 9. According to the plane mirror imaging principle, the points of the real array and the virtual array are symmetrical about the wall, and the virtual array is at point $(0, 2d_0)$. The two paths from the source to the real array and the virtual array can be formulated as:

$$\begin{aligned} y &= \tan(\theta_1 + \alpha)x, && \text{(LOS)} \\ y &= \tan(\theta_2 - \alpha)x + 2d_0, && \text{(ECHO)} \end{aligned} \tag{11}$$

where $\alpha$ is the array's orientation to the wall. Therefore, the point of intersection of these two lines is the source location. Note that before reverse ray-tracing, the distance $d_0$ and the orientation $\alpha$ need to be calibrated. We cover the calibration in § 8.1.

# 8 PRACTICAL ISSUES

## 8.1 Self-Calibration

To obtain geometric knowledge of the wall and finally localize sources, we need to know two parameters marked in Fig. 10 (a): the distance $d_0$ between the array and the wall, and the array's orientation $\alpha$ to the nearby wall.

We notice that besides the microphone array, the smart speaker, of course, has a speaker, which means the smart speaker can actively transmit probing signals to sense the environment. Specifically, we choose a sweep signal, chirp, to probe the wall. After transmitting the chirp, the smart speaker detects the arrival time for the echo of the chirp to each microphone. Therefore, the smart speaker can estimate the distance $d_0$ to the wall, based on the time taken for the chirp to bounce back to the smart speaker. Meanwhile, due to the orientation $\alpha$ , the distance $d_0$ measured by each microphone shall be different. These differences capture the orientation $\alpha$ . We then calculate the orientation $\alpha$ based on these differences.

## 8.2 Corner Case: Corner

A smart speaker might be placed at a corner, as shown in Fig. 10 (b). In this case, they will be two nearby walls and thus two main wall reflections. In other words, there will be three arriving paths LOS, ECHO0, and ECHO1 for each source. Similarly, we take two steps to localize sources: (1) estimate the DoAs of each path of each source, and (2) recognize each DoA (source and path type).

For DoA estimation, the method introduced in § 5 can be directly applied here. This is because the DoA estimation method is path-wise. As long as the path is directional, the resulting pure peaks' redundancy and coherency still exist, which can then be exploited to filter and refine DoAs.

For DoA recognition, *Symphony* now needs to recognize three paths: (1) identify which three DoAs belong to the same source. and (2) distinguish their path types, which is LOS or ECHO0 or ECHO1. Because the ECHO0 and ECHO1 paths are both the delayed versions of the LOS path, the inter-path coherence still exists. We can leverage hybrid peaks to recognize same-source DoAs (§ 6.1). To distinguish path types, we find a new criterion in the corner case: $\theta_2$ is always less than $\pi/2$ while $\theta_3$ is always larger than $\pi/2$. Based on this new criterion and the intersection criterion (§ 6.2), *Symphony* has sufficient constraints to distinguish three path types.

Finally, we point out that the Self-Calibration (§ 8.1) can also be used to detect whether the smart speaker is at a corner, and measure relative positions to two nearby walls: If two distinct echos of chirp are observed, the smart speaker then knows two walls are next to itself. Basically, by measuring the time of flight and time difference, the distance and the orientation to either wall can be derived.

## 8.3 How to determine the number of sources?

In fact, *Symphony* does not explicitly guarantee that a certain number of sources will be localized, instead provides a "best-effort" service. The whole localization procedure does not regulate that a certain number of sources *Symphony* must find. Instead, *Symphony* finds pure-peak combinations as many as possible, and thus localizes sources as many as possible. This means we implicitly determine the number of sources. Such design is actually based on our realization that in multi-source scenarios, determining the number of sources is not a simple counting problem. This is because some sources can be so close that their DoAs cannot be distinguished by a commercial array, or because a source may be so weak that its correlation peaks are buried below the noise floor and then its DoAs cannot be estimated.

# 9 EVALUATION

## 9.1 Implementation

**Hardware.** As shown in Fig. 11, We built a prototype of *Symphony* using commercial off-the-shelf microphone arrays with two different layouts: Seeed Studio ReSpeaker 4-mic linear array [33] and ReSpeaker 6-mic circular array [34]. These two layouts are widely used in many popular smart speakers such as Amazon Echo and Alibaba Tmall Genie. The inter-distance of two adjacent microphones is 5 cm for both the 4-mic linear array and the 6-mic circular array. The speed of sound is assumed as 343 m/s. The sampling rate is set to 48 KHz, which covers the whole audible frequency range. Each array is setting on top of a Raspberry Pi 4 Model B.

**Software and Algorithm.** We use the classical GCC method, GCC-PHAT [23], to calculate CCF, which whitens the microphone signals to equally emphasize all frequencies. The computational efficiency of *Symphony* is bottlenecked by the calculation of CCF. To accelerate the computation of GCC-PHAT, Fast Fourier Transform (FFT) is used. Meanwhile, zero padding in the frequency domain and interpolation are also applied to reduce the discretization error after performing FFT. To evaluate the computational efficiency of *Symphony* and prove that the localization overhead is affordable for resource-limited devices, We implement it in two computing platforms in Python: (1) using the local computing resource of a Raspberry Pi to localize the sources. (2) streaming audio samples to a more powerful device, a laptop, over WiFi, and the laptop executes the codes to localize sources.

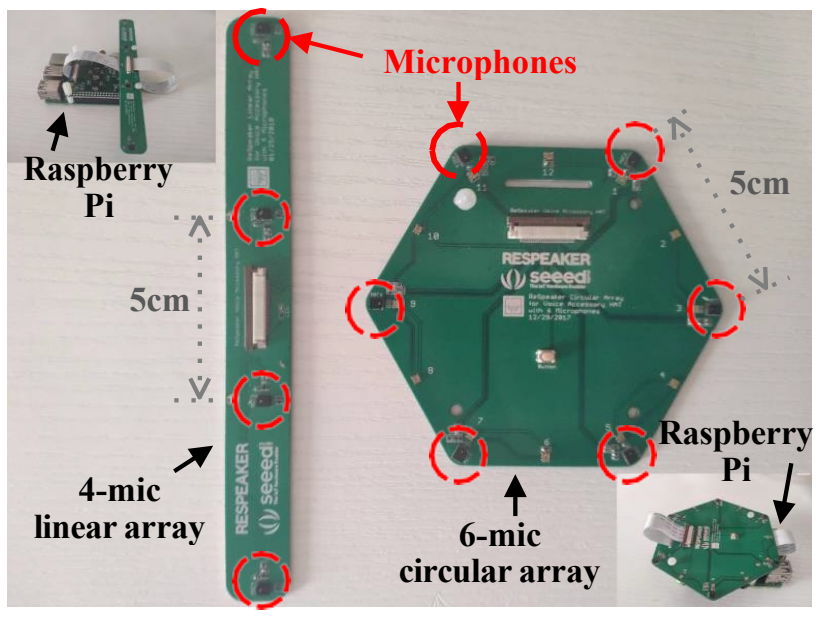

**Figure 11: Commercial off-the-shelf microphone arrays: 4-mic linear array and 6-mic circular array.**

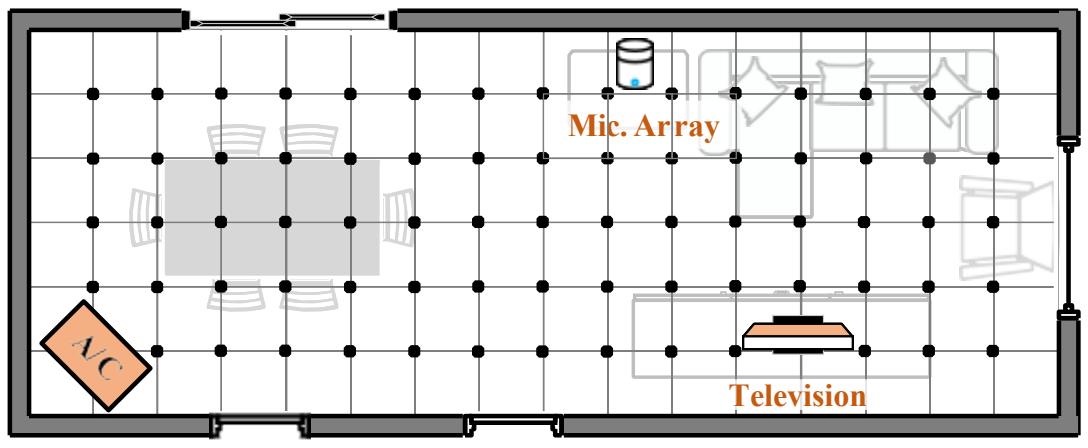

**Figure 12: The living room (8m x 3m).**

## 9.2 Experimental Methodology

In the experiments, four volunteers (including both men and women) are recruited to record different voice commands. The durations of these voice commands are between 0.2s and 10s. We use a portable wireless speaker as the acoustic source, which plays the recorded voice commands at different positions in the room. The sound volume of the speaker is set at around 60dB.

We compare *Symphony* with VoLoc [35], a state-of-the-art approach for single-source localization using a single array. We implement VoLoc on both 6-mic and 4-mic arrays. For fairness, the sampling rates of VoLoc and *Symphony* are both set to 48 kHz.

We conduct multiple experiments to evaluate the performance of *Symphony*. In the following subsections, we first present the localization result in the multi-source scenario (§ 9.3). Then, we compare the performance of Symphony with that of VoLoc in the single-source scenario, under both clean and noisy conditions (§ 9.4). We further present benchmarks, including computational efficiency (§ 9.5) and DoA estimation accuracy (§ 9.6). In addition to the above evaluation, we also conduct a case study to show the significance of *Symphony* in the safeguard application (§ 9.7).

## 9.3 Localization in the Multi-Source Scenario

*9.3.1 Overall Performance.* We first evaluate *Symphony*'s localization accuracy in the multi-source scenario. We conduct the experiment in a living room of 8m x 3m (Fig. 12). A 4-mic array is placed 0.35m away from the wall. The television is powered on and kept making sound (about 65dB), while the air conditioner is powered off. We use the portable speaker as the source and deploy it at different positions in the room, denoted by the black dots in Fig. 12. So we have two sources in this experiment. Fig. 13 shows the localization error of the speaker. Meanwhile, Fig. 14 plots the localization error of the television. From the results, we can see that:

- *Symphony*'s localization error tends to increase as the distance between the array and the source increases. According to the aforementioned far-field effect, due to the limited resolution of the hardware, the minimum measurable change in DoA is fixed.
- There is a blind region of localization, as shown in Fig. 13: When the connected line between the source and the array is perpendicular to the wall, *Symphony*'s localization performance is quite poor. To better understand this, let's refer to Fig. 9. The perpendicular case in Fig. 13 is equivalent to the case where the real array, the virtual array, and the source are col-linear. In this case, the slope of LOS is equal to that of ECHO, i.e., $\theta_1 = \theta_2$. It's impossible to know this source location using reverse ray-tracing. In other words, any source close to the negative Y-axis cannot be localized. In our implementation, *Symphony* simply outputs the array location when detecting $\theta_1 \approx \theta_2$.
- The localization performance degrades when two sources are close. This is because when the sources are close, their DoAs of the LOS and the ECHO paths are also close and indistinguishable, thus confusing *Symphony* during the DoA estimation.

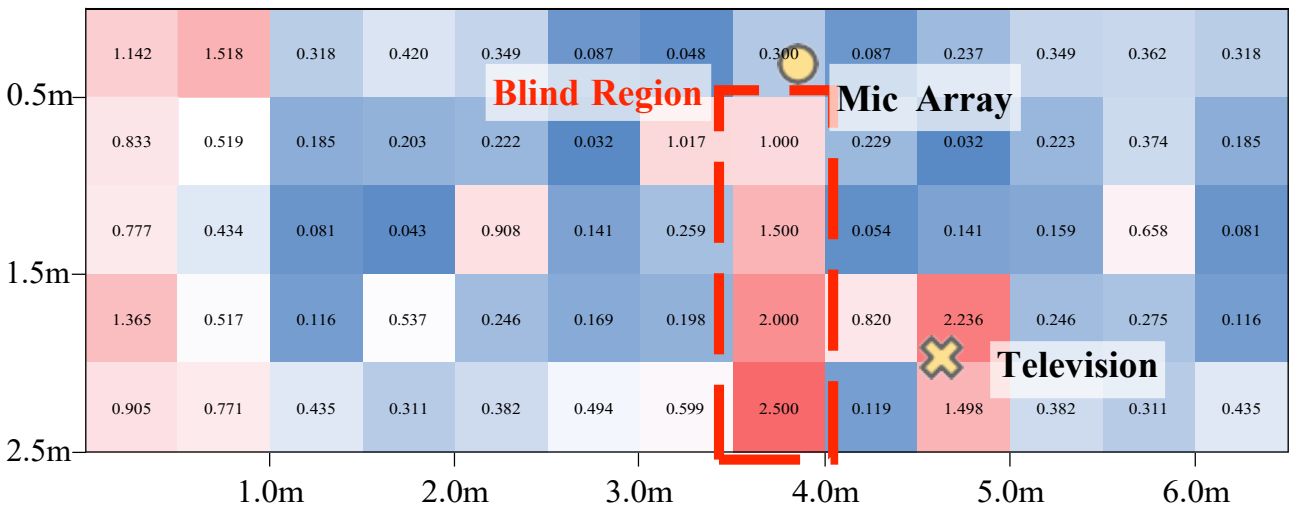

**Figure 13: Heatmap of *Symphony*'s localization error of the voice commands in the living room. The yellow circle and the yellow cross denote the array and the television.**

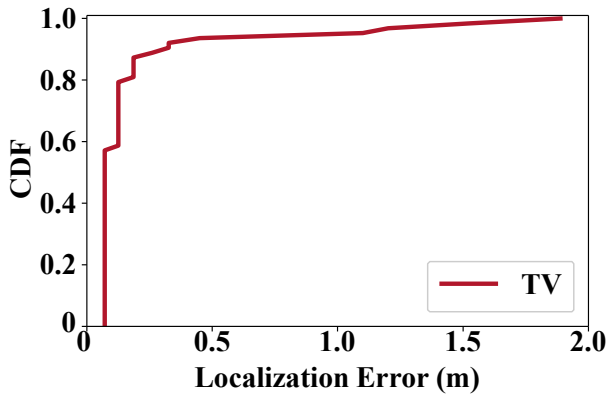

**Figure 14: Localization error of the television.**

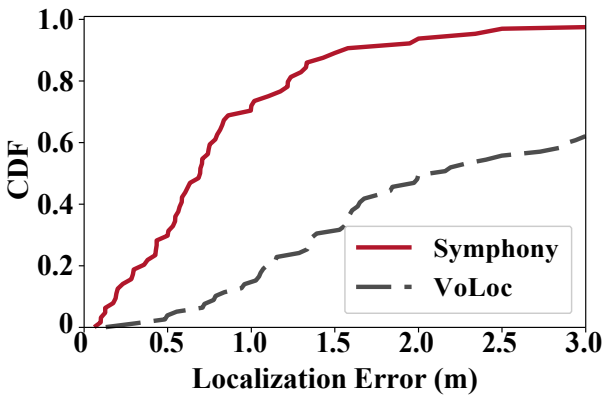

**Figure 15: Localization error of the voice (6-mic).**

9.3.2 *Comparison.* Next, in the same multi-source scenario, we compare the localization performance of *Symphony* and VoLoc. Note that VoLoc can only localize a single source. VoLoc's localization performance in this case is actually quite uncertain. For purpose of comparison, we plot *Symphony*'s localization result of the speaker and the localization result of VoLoc together in Fig. 15. This time, we use the 6-mic array for localization. The array is placed 0.4m away from the wall. As we can see, *Symphony* achieves a satisfactory accuracy, with a median error of 0.694 m, while VoLoc suffers dramatic performance degradation, with a median error of 2.161 m. *Symphony* reduces the localization error by 68%, compared with VoLoc. VoLoc's poor performance is mainly due to the interference between the two acoustic sources (the speaker and the television). There might be an option to alleviate this problem of VoLoc, by using a band-pass filter to separate one source from the other, which have different frequency bands. Nevertheless, a acoustic source like the television shares a wide frequency band with human voice. Such an option may be infeasible in practice.

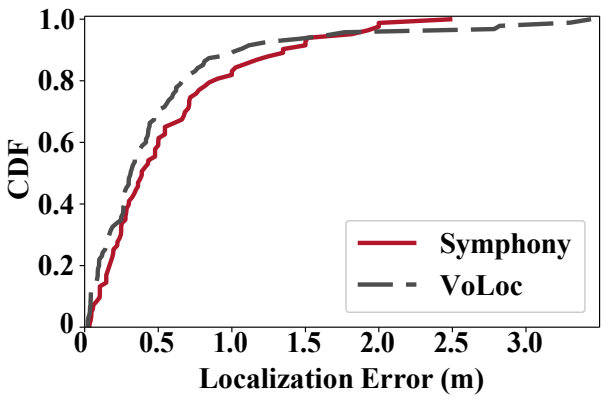

**Figure 16: Localization error on ideal conditions (4-mic).**

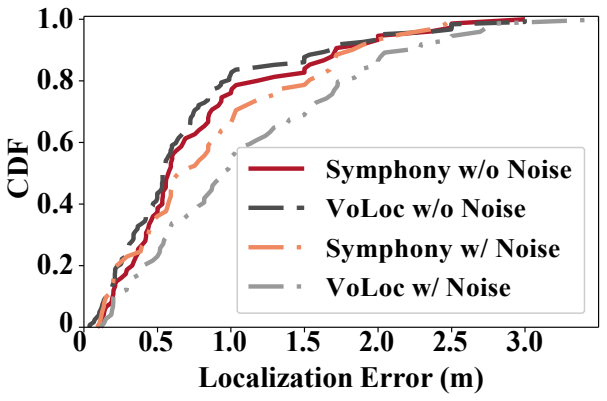

**Figure 17: Localization error w/ or w/o noise (6-mic).**

## 9.4 Localization in the Single-Source Scenario

We compare the localization accuracy of *Symphony* with VoLoc in the single-source scenario, under both clean and noisy conditions. Owing to the ability to distinguish multiple acoustic sources, we expect that *Symphony* has better performance than VoLoc in the noisy environment. The experiments are conducted in the living room (Fig. 12). We play the voice commands via a portable speaker at different positions.

**Localization under clean condition.** We place the 4-mic array 0.4m away from the wall. We conduct experiments in the night, when it is very quiet in the living room. The background noise is lower than 20 dB. Fig. 16 shows the localization error of *Symphony* and VoLoc. The median error of VoLoc is 0.314 m, which is slightly better than that of *Symphony*, 0.387 m. The slight gap in performance is because VoLoc uses a fine-grained but exhausting searching method to localize the source, which produces more accurate results in the ideal case.

**Localization under noisy condition.** We place the 6-mic array 0.4m away from the wall. We conduct the experiment in the daytime. The volume of background noise is around 40-46 dB. In addition, the air conditioner is kept on, producing noise at about 38-42 dB. The television is powered off. For completeness of the result, we also plot the localization result of *Symphony* and VoLoc using the 6-mic array under clean condition. In the noisy environment, *Symphony* only suffers slight performance degradation. The median error increases from 0.578m to 0.662m. The performance of VoLoc suffers significant degradation. The median error increases from 0.536m to 0.937m. VoLoc is susceptible to the noise because it only uses a short time window of the received signals to search for DoAs. So its performance is likely to be affected by the low signal-to-noise ratio. *Symphony* uses the whole voice signals to obtain the CCFs, which is robust against the noise.

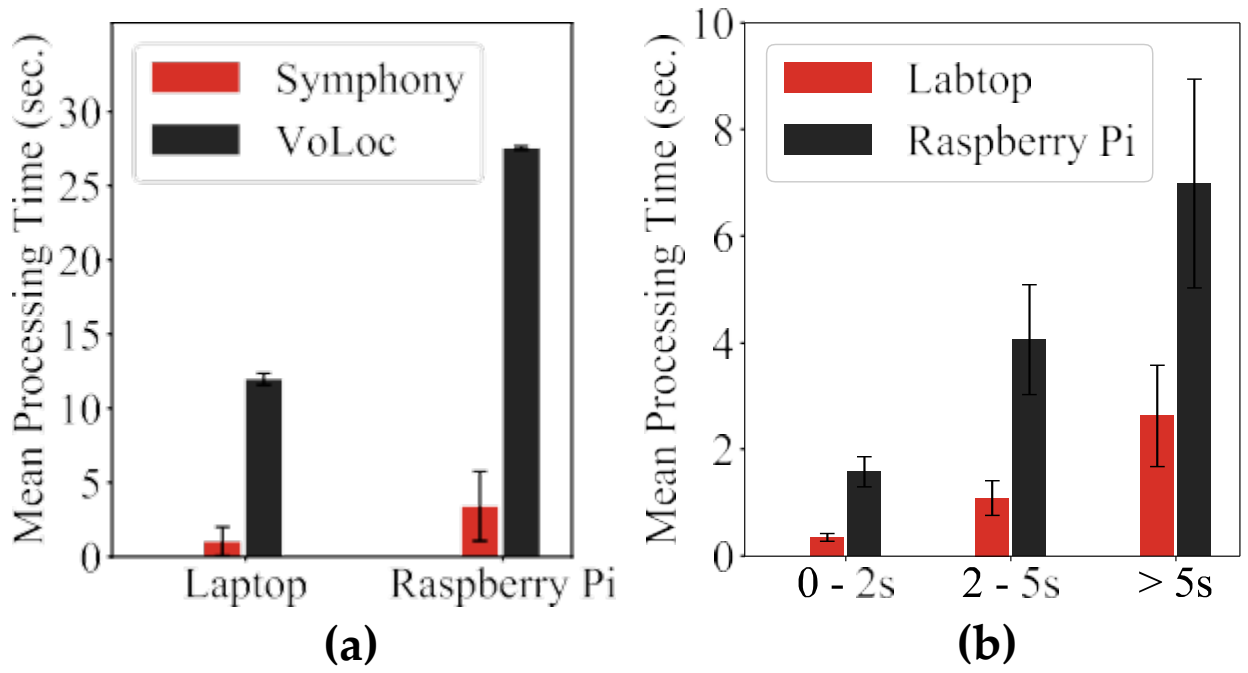

**Figure 18: (a) Comparison on processing time. (b) Symphony's processing time across different voice durations.**

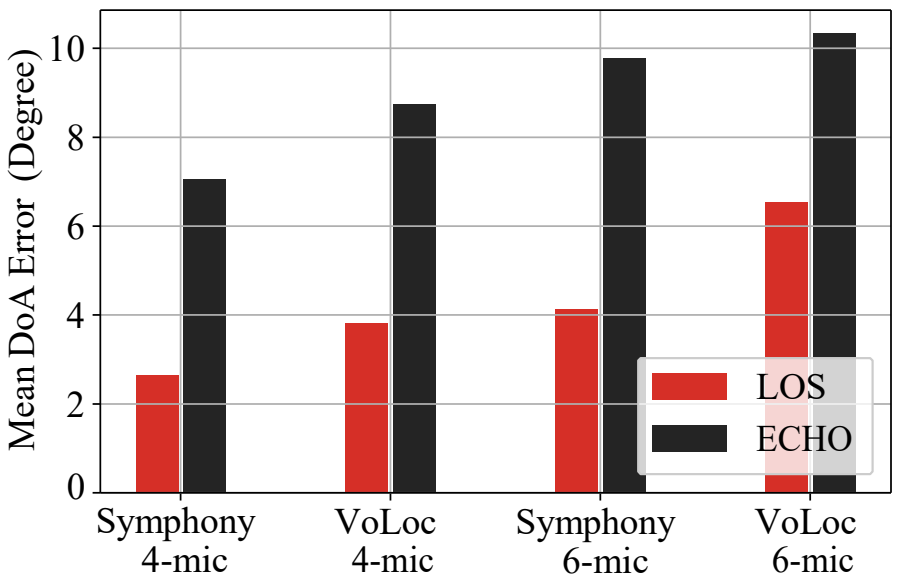

**Figure 19: DoA estimation error of *Symphony* and the baseline using 4-mic linear array and 6-mic circular array.**

Moreover, by comparing the results under the clean condition in Fig. 16 and Fig. 17, we find that for both *Symphony* and VoLoc, the 4-mic array version performs better than the 6-mic array version. This is because the 4-mic array has a larger microphone inter-distance than the 6-mic array, leading to a finer spatial resolution [40].

## 9.5 Computational Efficiency

*Symphony* can efficiently process audio samples in real-time. We evaluate the gain in reducing processing time that *Symphony* delivers over VoLoc. We run *Symphony* and VoLoc on two devices — the laptop and the Raspberry Pi, and measure the processing time needed to localize the source.

Fig. 18 (a) compares the mean and the standard deviation of processing time between *Symphony* and the baseline. As we can see, the average time consumption of *Symphony* for the laptop and the Raspberry Pi are 1.4 seconds and 3.4 seconds, respectively. Compared to the baseline, *Symphony* significantly reduces the processing time, and is 9x and 8x faster than the baseline on the laptop and the raspberry pi, respectively. This is because VoLoc models the localization problem as a highly non-convex optimization problem, and resorts to the brute-force searching to localize the source, which is quite time-consuming. On the contrary, *Symphony* fully takes advantage of the sparsity of peaks to avoid sample-wise searching, as well as the geometry of the array to quickly narrow down to pure-peaks combinations with high possibility. We believe the time overhead of *Symphony* can be further reduced such as using multiple threads to compute CCFs simultaneously, and are totally affordable for low-cost devices like Raspberry Pi.

We notice that in Fig. 18 (a), the standard deviations of VoLoc's are smaller than *Symphony*, which means VoLoc's processing time is relatively static. This is because although VoLoc searches DoAs over a large space, the size of the searching space is fixed. On the other hand, the large standard deviations of *Symphony* results mostly from the fluctuation of CCF computation: voice commands with different durations will lead to different time consumption of CCF computation. To study this factor, we categorize all our commands into three groups based on duration: 0 2s, 2 5s, and > 5s. Fig. 18 (b) shows *Symphony*'s processing time over three categories. We observe that as the duration decreases, the processing time also decreases. When the duration is less than 2s, the processing time is only 0.35s and 1.58s for the laptop and the Raspberry Pi.

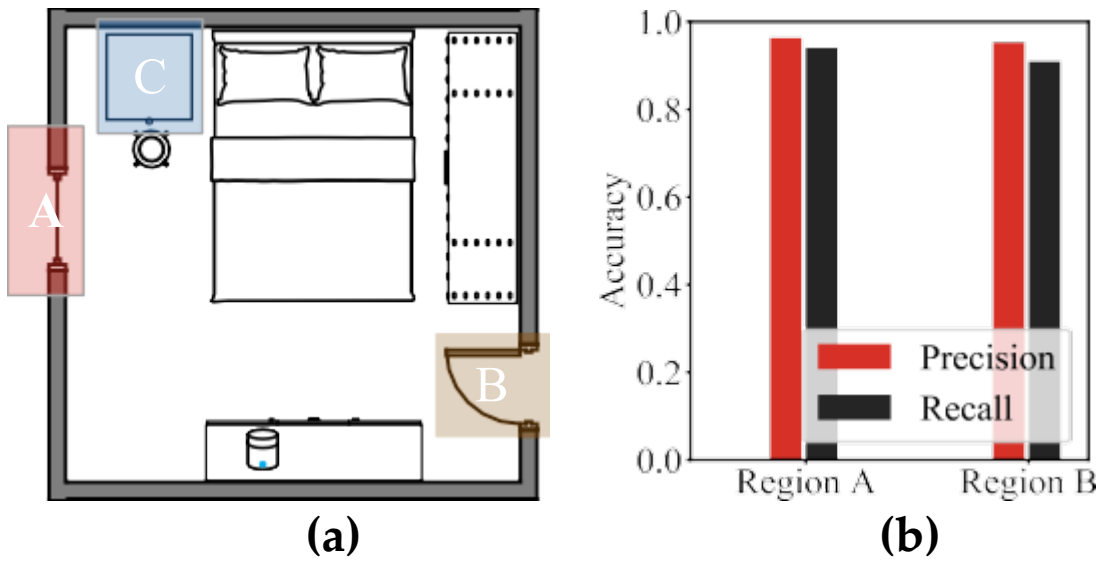

**Figure 20: (a) A bedroom (3m x 3m). The smart speaker detects whether there are abnormal sounds appearing on two regions A and B, the window and the door. The sound is allowed to appear in the region C, the bedside table. (b) Precision and Recall of abnormal sound detection.**

## 9.6 DoA Estimation Accuracy

Here, we evaluate the accuracy of DoA estimation, which directly determines the accuracy of localization. Fig. 19 compares *Symphony* and the baseline on DoA estimation error in the environment with ambient noise. As we can see, the DoA estimation of the LOS path, as expected, is more accurate than the ECHO path. We also observe that the 4-mic array obtains a finer DoA resolution than the 6-mic array. This is because, again, the 4-mic array has a larger microphone interval than the 6-mic array. Meanwhile, in the noise environment, *Symphony* achieves a better DoA estimation than the baseline, which is consistent with the results shown in Fig. 17.

## 9.7 Case Study: Safeguard

In this case, we envision that *Symphony* is able to monitor abnormal sounds for certain places in our homes. With this ability, the smart speaker can warn the hosts that something unusual happened, e.g., the window is broken, or the room is invaded. In the experiment, we monitor three regions in a bedroom, as shown in Fig. 20 (a), including two abnormal areas and one normal area. The array is placed 0.5m away from the wall. The abnormal areas are not supposed to have sounds during sleeping while the normal area is allowed to have. We then make sounds in different places in this room, and use *Symphony* to monitor the room. We keep the phone ringing on the bedside table. Fig. 20 (b) shows the precision and the recall of detection. As we can see, *Symphony* has high accuracy in identifying abnormal acoustic sources with 96.4% precision and 95.3% recall at region A, and 0.94% precision and 0.91% recall at region B. *Symphony* demonstrates the feasibility of safeguard.

## 10 RELATED WORK

**Localization by Exploiting Room Information.** The work closest to our scenario is VoLoc [35]. VoLoc observes an interesting application-specific opportunity that smart speakers are typically near a wall to connect to power outlets. Based on such observation, VoLoc demonstrates the feasibility of using a single (small) device to localize a single acoustic source by inferring the nearby wall reflection. In [2], not only the nearby wall, the whole room information is measured by a Kinect depth sensor, and then used to localize a clapping sound. [8, 24] measure the shape of the room and localize the source via transmitting and receiving previously known signals. This method can not be applied in our case where the sources are unknown. Meanwhile, many works do not explicitly measure room information, instead incorporate room information into fingerprints built on top of RSSI, CSI, electromagnetic or geomagnetic field, and visible light [26, 28, 36, 37, 46, 47, 49]. However, these approaches require laborious work to build fingerprint dataset.

**Concurrent Localization.** The most successful concurrent Localization system, perhaps, is the Global Positioning System (GPS), which supports billions of clients localizing themselves concurrently. Such nearly unlimited scalability is, partially, because channels are utilized by only the anchors (i.e., satellites). the clients only need to listen without occupying channels. This design is rather appealing. Chorus [6] and SnapLoc [15] introduce a completely different scheme in which UWB tags do not transmit at all and localize themselves only by measuring channel impulse response (CIR), which significantly improves scalability compared with standard UWB two-way ranging (TWR) [16]. However, when clients actively join the procedures of localization, the scalability decreases dramatically, because clients should transmit in different time slots to avoid collisions. Many methods are designed to tolerate collisions from multiple clients, which enable parallel transmission/sensing and pave the way for concurrent localization [1, 12, 17, 20–22, 29]. For example, LF-Backscatter [17], BiGroup [29], FlipTracer [22], HUBBLE [21], and Fireworks [20] decode the collided packets by leveraging the spatial and/or temporal patterns of signals' combined states. Inspired by these works, *Symphony* tolerates the collisions between sources and measures their DoAs from collided signals. Recently, [39] tries to localize multiple sources via a deep learning method. However, This method can not be easily generalized to new environments that the training dataset does not cover.

**Localization with Multiple Arrays.** Distributed microphone / antenna arrays have been used to localize various sources, including WiFi clients [45], RFID tags [40], birds [5], and bumblebees [19]. ArrayTrack [45] aggregates DoA information computed by several WiFi APs, and then estimates the WiFi client's location. RF-IDraw [40] uses two RFID readers, each connected with four antennas, to localize tags. Living IoT [19] enables insect-borne sensors to localize bumblebees using two arrays.

**DoA Estimation.** *Symphony* uses a popular method, GCC-PHAT [23], to compute cross-correlation function (CCF). In fact, *Symphony* might be extended to other DoA algorithms like MUSIC [32] and ESPRIT [18], because the underlying problem is the same: the ambiguity of peaks, no matter peaks are from the CCF (GCC-PHAT), or from the pseudo-spectrum (MUSIC and ESPRIT). Our propagation model is still valid, and our insights in the geometric redundancy and the coherence of received signals also exist, which can be exploited to solve the ambiguity.

## 11 DISCUSSION

- **3D Localization.** *Symphony* virtually contains two microphone array, the real one and the virtual one, and thus only supports localization in 2D.
- **Non-Line-of Sight.** A source is localized as the intersection of the LOS path and the ECHO path. When the LOS path is blocked, *Symphony* will fail to localize the source.
- **The Array-Wall Distance.** If the array is far away from the wall, some problems will arise: (1) The ECHO path will experience more propagation attenuation and its strength would not be comparable to the LOS path. (2) The ECHO path is unlikely to be the second arriving path. For example, the path reflected by the wall very close to the source may be shorter than ECHO, and may be stronger than ECHO. On the other hand, if the array is too close to the wall, *Symphony* will also perform poorly. This is because in this case the real array becomes close to the virtual array, and the far-field effect introduced in § 2 will happen again. According to our experience, the appropriate distance between the array and the wall is between 0.25m and 0.6m.
- **Number of Microphones.** *Symphony* requires at least 3 microphones to explore geometric redundancy in multiple pairs. For some smart speakers which contain only two microphones (e.g., Google Home), *Symphony* might not be applied to them.
- **Moving Objects.** *Symphony* assumes that sources are static when sources are active. Localizing moving objects might be another problem, which is out of the scope of this paper. One possible solution is to use a small time window of signals, during which sources can be viewed as static, to trace sources.

## 12 CONCLUSION

In this paper, we propose *Symphony*, the first approach to concurrently localize multiple acoustic sources with a single microphone array. *Symphony* enables the smart speaker to localize users under interference, and then improves the cognitive ability. We believe *Symphony* might further open up a whole new class of applications in location-aware services and user interaction interface.

## ACKNOWLEDGMENTS

We thank the anonymous shepherd and reviewers for their comments and feedback. This work is supported in part by National Key R&D Program of China No. 2017YFB1003000, National Science Fund of China No. 61772306, the Smart Xingfu Lindai Project, and the R&D Project of Key Core Technology and Generic Technology in Shanxi Province (2020XXX007).